\documentclass[preprint,prd,aps,showpacs,showkeys,nofootinbib]{revtex4-1}
\usepackage{amssymb}
\usepackage{amsmath}
\usepackage{graphicx}
\usepackage{subfigure}
\usepackage{float}
\usepackage{mathrsfs}
\usepackage{amsmath}
\newcommand{\ba}{\begin{eqnarray}}
\newcommand{\ea}{\end{eqnarray}}

\begin{document}
\title{Taiji Program: Gravitational-Wave Sources}

\author{Wen-Hong Ruan}
\email{ruanwenhong@itp.ac.cn}
\affiliation{CAS Key Laboratory of Theoretical Physics, Institute of
Theoretical Physics, Chinese Academy of Sciences, P.O. Box 2735,
Beijing 100190, China}
\affiliation{School of Physical Sciences,
University of Chinese Academy of Sciences,
No. 19A Yuquan Road, Beijing 100049, China}
\author{Zong-Kuan Guo}
\email{guozk@itp.ac.cn}
\affiliation{CAS Key Laboratory of Theoretical Physics, Institute of
Theoretical Physics, Chinese Academy of Sciences, P.O. Box 2735,
Beijing 100190, China}
\affiliation{School of Physical Sciences,
University of Chinese Academy of Sciences,
No. 19A Yuquan Road, Beijing 100049, China}

\author{Rong-Gen Cai}
\email{cairg@itp.ac.cn}
\affiliation{CAS Key Laboratory of Theoretical Physics, Institute of
Theoretical Physics, Chinese Academy of Sciences, P.O. Box 2735,
Beijing 100190, China}
\affiliation{School of Physical Sciences,
University of Chinese Academy of Sciences,
No. 19A Yuquan Road, Beijing 100049, China}

\author{Yuan-Zhong Zhang}
\email{zyz@itp.ac.cn}
\affiliation{CAS Key Laboratory of Theoretical Physics, Institute of
Theoretical Physics, Chinese Academy of Sciences, P.O. Box 2735,
Beijing 100190, China}
\affiliation{School of Physical Sciences,
University of Chinese Academy of Sciences,
No. 19A Yuquan Road, Beijing 100049, China}

\begin{abstract}
We review potential low-frequency gravitational-wave sources, which are expected to be detected by Taiji,
a Chinese space-based gravitational-wave detector,
estimate the detection rates of these gravitational-wave sources
and present the parameter estimation of massive black hole binaries.
\end{abstract}


\maketitle

\section{Introduction}
On 11 February 2016 it is announced that on 14 September 2015 the two detectors of the Laser Interferometer
Gravitational-Wave Observatory (LIGO) simultaneously observed a transient gravitational wave (GW) signal
from the coalescence of a stellar mass black hole (BH) binary, which is named GW150914~\cite{Abbott:2016blz}.
This is the first direct detection of GWs and the first observation of a binary BH merger.
This great discovery confirms the existence of GWs
first predicted by Einstein's general theory of relativity in 1916.
So far several GW events have been detected, especially GW170817, a binary neutron star inspiral~\cite{TheLIGOScientific:2017qsa}.
The detections of GWs together with their electromagnetic counterparts
indicate the coming of the era of GW astronomy and GW cosmology.
Hence, GWs open a new window to explore our Universe.

Compared to GWs in the frequency range of $10-10^3$ Hz,
GWs at low frequencies from $0.1$ mHz to $1.0$ Hz are expected to carry an enormous amount of information
on galaxy formation, galactic nuclei, the Milky Way and the early Universe~\cite{Seoane:2013qna}.
Such new GW sources, which are expected to be detected by future space-based interferometers,
can be divided into two categories.
One is compact binary coalescences including massive black hole binaries (MBHBs),
extreme/intermediate mass ratio inspirals (EMRIs/IMRIs)
and compact binaries in the Milk Way.
The other is stochastic gravitational-wave backgrounds (SGWBs) from inflation, reheating/preheating after inflation,
and first-order phase transition in the early Universe~\cite{Cai:2017cbj}.
In the present article, we first review these GW sources,
then estimate the detection rates of these GW sources by
Taiji, a Chinese space-based GW detector~\cite{Hu:2017cbj},
and present the parameter estimation of MBHBs.

\section{GWs from compact binary coalescences}
\subsection{Massive black hole binaries}
Substantial observations support the existence of massive BHs with masses from $10^5 M_\odot$ up to $10^{10} M_\odot$
in the centers of almost all galaxies.
However, the origin of the massive BHs is still an open question.
These massive BHs are thought to grow from smaller ``seed'' BHs which are created in the early Universe.
Such seed BHs may be the remnants of Population III stars with masses $\sim 10^2 M_\odot$~\cite{Madau:2001sc}
or may form due to the collapse of proto-galaxies with masses $\sim 10^4 - 10^6 M_\odot$~\cite{Koushiappas:2003zn} at high redshifts $z \sim 15-20$.
In both cases the initial mass functions for the seeds have large uncertainties.

The coalescences of MBHBs are the strongest GW sources in the frequency range $\sim 10^{-8} - 1$ Hz,
which have been one of the main targets of space-based gravitational wave observatories~\cite{Hughes:2001ya,Barausse:2014oca,Klein:2015hvg}.
The GW signals from MBHBs contain three distinguished parts: inspiral, merger and ringdown,
which are expected to be detected with extremely high signal-to-noise ratios based on matched filtering~\cite{Cao:2017ndf}.
More numerical relativity study is needed for the template construction~\cite{Cao:2011fu,Cao:2008wn},
especially for the mass ratio more than $1:20$ binary BH systems.
Since massive BHs are embedded in dense galactic nuclei, surrounded by gas and stars,
the interaction of MBHBs with their environment may provide a possibility to simultaneously observe GWs and electromagnetic counterparts.
Once these host galaxies are identified with the help of the electromagnetic counterparts,
MBHBs may be used as ``standard sirens'' to probe the expansion history of the Universe at high redshifts.
Moreover, GWs from MBHBs carry precious information on the co-evolution between massive BHs and their host galaxies
and on the dynamics of gas and accretion onto massive BHs,
which is thus crucial to our understanding of the growth of structure.

\subsection{Extreme/intermediate mass ratio inspirals}
It is known that there are $\sim 10^7-10^8$ stars including compact objects
(either white dwarfs, neutron stars or stellar mass BHs)
surrounding the massive BH in the center region of galaxies with a size of a few pc~\cite{AmaroSeoane:2007aw}.
Due to high stellar densities in this region collisional effects come into play.
A compact object may directly be plunged into the masive BH
or inspiral gradually around the massive BH due to the emission of GWs prior to plunge.
The latter process is the so-called EMRI/IMRI
with mass ratios $<10^{-4}$ or $\sim 10^{-4}-10^{-2}$~\cite{AmaroSeoane:2007aw}.

EMRIs/IMRIs are among the most interesting GW sources
that are expected to be detected by space-based interferometers.
Since the EMRI/IMRI waveform carries an enormous amount of information,
the observation of such a GW signal gives a new way to test general relativity
and the structure of spacetime surrounding massive BHs,
as well as allows us to understand the dynamics of compact objects in galactic nuclei~\cite{Barack:2003fp}
and to measure the equation of states of white dwarfs~\cite{Han:2017kre}.
However, the large parameter space of EMRIs/IMRIs makes a number of templates
required by matched filtering computationally prohibitive.
Therefore, the EMRI/IMRI search will rely on fast but accurate approximations to EMRI/IMRI waveforms.
The accuracy of templates is decided by two issues: the physical model itself and the computation precision.
A hybrid method, which combines effective-one-body dynamics and Teukolsky equation,
may improve the accuracy of numerical calculation of EMRI/IMRI waveforms
by including the mass-ratio corrections and eccentricity~\cite{Han:2011qz,Han:2014ana}.

\subsection{Compact binaries in the Milk Way}
The majority of stars in the Milk Way are thought to exist in binary systems,
about half of which are compact binaries with white dwarfs, neutron stars and stellar mass BHs.
The population of compact binaries is calculated in~\cite{Nelemans:2001hp}.
The number of double white dwarfs in the Galactic disk is $1.5 \times 10^8$,
which vastly outnumber all other binaries with compact objects in the Galactic disk.

Compact binaries in the Milk Way are expected to be
among the most numerous and loudest signals detected by space-based interferometers.
The strain amplitude $h$ and the change of frequency $f$ of the GW signal are given by~\cite{Postnov:2014tza}
\ba
h &\propto& M_c^{5/3}f^{2/3}d_L^{-1}, \\
\dot{f} &\propto& M_c^{5/3}f^{11/3}, \\
\ddot{f} &\propto& \dot{f} f^{-1},
\ea
where $M_c$ is the chirp mass and $d_L$ is the luminosity distance to the source.
It is shown that the GW signals from the Galactic disk are dominated
by double white dwarfs and that their number is so large that
they form the GW confusion limited signal at frequencies $f<2$ mHz.
Such the confusion signal is dubbed a foreground relative to backgrounds produced in the early Universe.
The spectral shape of the foreground contains information about the homogeneity of the sample,
which provides information on the distribution of the sources in the Galaxy disk.
Above a certain limiting frequency (somewhere between $\sim 1-10$ mHz) several thousand of double white dwarfs and few
tens of binaries containing neutron stars will be resolved,
fewer than 50 of which are currently known.
The study of the GW signals from these resolved compact binaries would help intercept
information on the star formation history and structure of the Galaxy,
the chemical composition of white dwarfs, formation channels of these binaries,
and evolution of their progenitors~\cite{Yu:2011mw,Yu:2012tw,Yu:2015yaa}.

\section{GWs from the early Universe}
\subsection{Inflation}
To solve some theoretical problems in big-bang cosmology such as the horizon and flatness problems, inflationary
scenario was proposed and developed~\cite{Guth:1980zm,Albrecht:1982wi,Linde:1981mu},
in which a period of accelerated expansion of the Universe happened at early times.
Inflation not only predicts the primordial scalar perturbations, which provide a natural way
to generating the anisotropies of the CMB radiation and the initial tiny seeds of the large-scale
structure observed today in the Universe, but also generates primordial tensor perturbations (called primordial GWs)
which result in B-mode polarization of the CMB anisotropies.
Although such a SGWB has not been observed yet, its detection would open a new window
to understanding the early Universe physics and thus the origin and evolution of the Universe.

GWs produced during inflation are described by a transverse-traceless gauge-invariant tensor perturbation, $h_{ij}$,
in a Friedman-Robertson-Walker metric, which satisfies at first order in perturbation theory
\ba
h_{ij}^{\prime\prime}+2{\cal{H}}h_{ij}^{\prime}-\nabla^2 h_{ij}=0,
\label{eq:inflation}
\ea
where the prime denotes the derivative with respect to the conformal time $\tau$,
${\cal{H}}$ is the Hubble parameter in $\tau$ and
$h_{ij}$ satisfy $\partial^i h_{ij} = 0$ and $\delta^{ij} h_{ij} = 0$.
In the slow-roll approximation, Eq.~(\ref{eq:inflation}) can analytically be solved.
The Bunch-Davies vacuum condition in the asymptotic past is imposed
because the modes lie well inside the Hubble radius.
During inflation, quantum fluctuations are amplified and stretched,
and then nearly frozen on super-Hubble scales.
The single-field slow-roll inflation predicts a slightly red-tilted spectrum of the SGWB.
Moreover, there is a consistency relation $n_T=-r/8$ between the
tensor spectral index $n_T$ and the tensor-to-scalar ratio $r$.
In the presence of a Gauss-Bonnet coupling to the inflaton field,
the consistency relation is broken~\cite{Guo:2006ct,Satoh:2008ck,Guo:2009uk,Guo:2010jr,Jiang:2013gza},
which differs from the single-field slow-roll inflationary scenario.
The shape of the spectrum is characterized by the tensor spectral index
since the running of the spectral index is negligible to the lowest order in slow-roll parameters.
More general shapes beyond slow-roll may be reconstructed by using a binning method of a
cubic spline interpolation in a logarithmic wavenumber space~\cite{Guo:2011re,Guo:2011hy,Hu:2014aua}.

Since tensor perturbations begin to oscillate with the amplitude damped by a factor $a^{-1}$
after reentering the Hubble horizon at the epoch of radiation or matter dominance,
for the slightly red-tilted spectrum, the energy spectrum of the SGWB
in the frequency range of $10^{-10}-10^3$ Hz at present becomes too weak to be directly detected
by pulsar timing array experiments and laser interferometer experiments~\cite{Boyle:2005se}.
However, for a blue-tilted spectrum predicted by inflationary models
beyond slow-roll or caused by the source term such as particle production~\cite{Cook:2011hg} and spectator fields~\cite{Biagetti:2013kwa},
the direct detection might be possible for future space-based interferometers~\cite{Bartolo:2016ami}.

\subsection{Reheating/Preheating}
In the inflationary scenario, the inflaton field begins to oscillate
around the minimum of its potential after inflation.
Such coherent oscillations produce the elementary particles known to us and eventually reheat the Universe.
This process is called reheating~\cite{Albrecht:1982mp,Dolgov:1982th,Abbott:1982hn},
which sets the initial conditions of the hot Big Bang.
The coupling between the inflaton field and other fields is necessarily
tiny ensuring that reheating proceeds slowly.
If the initial amplitude of oscillations of the inflaton field is large enough,
preheating provides a more rapidly efficient mechanism for extracting energy from the inflation
field by parametric resonance~\cite{Kofman:1997yn}.
Such a process is so rapid that the produced particles are not in thermal equilibrium.
The preheating leads to large and time-dependent inhomogeneities of the
stress tensor that source a significant SGWB,
as pointed out by Khlebnikov and Tkachev in~\cite{Khlebnikov:1997di}.
So Eq.~\eqref{eq:inflation} becomes
\ba
h_{ij}^{\prime\prime}+2{\cal{H}}h_{ij}^{\prime}-\nabla^2 h_{ij} = \frac{2}{M_{pl}^2} \Pi_{ij}^{TT},
\label{eq:gwsource}
\ea
where $M_{pl}$ is the reduced Planck mass and
the source term $\Pi_{ij}^{TT}$ is the transverse-traceless projection of the anisotropic stress tensor $T_{ij}$.

Unlike GWs produced during inflation, they are generated and remain in the Hubble horizon until now.
Since their wavelengths are smaller than the Hubble radius at the time of GW production,
the peak frequency of this type of SGWB is typically of order more than $10^3$ Hz.
Detecting such a high-frequency SGWB is particularly challenging.
For example, for the $\phi^4$ and $\phi^2$ chaotic inflationary models,
lattice simulations show that preheating can lead to GWs with frequencies of around $10^6-10^8$ Hz
and peak power of $\Omega_\mathrm{GW}h^2 \sim 10^{-9}-10^{-11}$ at present~\cite{Easther:2006gt}.
It is found that the present peak frequency of such a SGWB is proportional to the energy scale of inflation~\cite{Easther:2006gt},
while the present amplitude of GWs is independent of the energy scale of inflation~\cite{Easther:2006vd}.
For hybrid inflation, since the energy scale of inflation can be chosen to range from GUT scales to TeV scales,
produced GWs cover a larger range of frequencies~\cite{GarciaBellido:2007dg}.

Oscillons can form after inflation with a family of potentials.
These oscillons are spatially-localized, oscillatory, long-lived field configurations.
In such a scenario, the Universe undergoes a transient matter-dominated phase.
In the models with a symmetric smooth potential~\cite{Zhou:2013tsa}
and an asymmetric smooth potential~\cite{Antusch:2016con},
oscillon formation can generate a SGWB with a characteristic structure in the energy spectrum.
For a cuspy potential, the cusp can trigger amplification of fluctuations of the inflaton itself
at the moment when $\phi(t)=0$, so that oscillons copiously form after inflation,
which sources a characteristic energy spectrum of GWs with double peaks~\cite{Liu:2017hua,Liu:2018rrt}.
The discovery of such a double-peak spectrum could test the underlying inflationary physics.
Similar to GWs produced during preheating,
the peak frequency of the energy spectrum depends on the energy scale of inflation.
It is expected to directly detect the signal by future ground-based or space-based interferometers.

\subsection{Phase transition}
First-order phase transitions are predicted in some scenarios beyond the Standard Model of particle physics.
If a strong first-order phase transition occurs in the early Universe,
many true vacuum bubbles are nucleated in a sea of the false vacuum.
These bubbles then rapidly expand and eventually collide with each other, which can generate a significant SGWB.
Therefore, the detection of such a SGWB provides a promising way to probe new physics beyond the Standard Model~\cite{Caprini:2015zlo}.

The strong first-order phase transition proceeds by the bubble nucleation, bubble expansion and bubble percolation.
In the first process, the false vacuum decays into the true vacuum via quantum tunnelling and
the true vacuum bubbles are nucleated~\cite{Coleman:1977py}.
Given the effective potential of the relevant scalar field, there are two key parameters controlling the GW signal.
One is the fraction $\beta/H_*$ where $\beta$ is the inverse time duration of the phase transition
and $H_*$ is the Hubble parameter at the nucleation temperature $T_*$.
The other is the strength $\alpha$, the ratio of the vacuum energy density released in the
transition to that of the radiation bath.
If the bubbles are smaller than a critical size their volume energy cannot overcome the surface tension and they disappear.
When the temperature drops below a critical temperature,
it becomes possible to nucleate bubbles that are larger than this critical size.
The bubble wall rapidly expands until approaching the speed of light.
Actually a thermal phase transition involves slower velocity of the bubble wall $v_w$
due to the friction term coupling the scalar field to the hot plasma full of relativistic particles.
The third process involves three sources of GWs,
namely, bubble collisions, sound waves in the plasma and magnetohydrodynamic turbulence in the plasma,
so that the predicted energy spectrum of the SGWB is
\ba
\Omega_\mathrm{GW}h^2=\Omega_{\phi}h^2+\Omega_\mathrm{sw}h^2+\Omega_\mathrm{turb}h^2.
\ea
These contribution expressions involve $\kappa_\phi$, $\kappa_v$ and $\kappa_\mathrm{turb}$,
the efficiency factors for conversion of the latent heat into kinetic energy of the wall,
bulk motion and magnetohydrodynamic turbulence, respectively.

For the strong first-order phase transition taking place at the electroweak epoch,
the peak frequency of the GW signal is typically in the milliHertz range,
just the most sensitive frequency of a space-based interferometer.
For a slow first-order phase transition, the signal peak shifts to lower frequencies~\cite{Cai:2017tmh,Cai:2018teh}.
A strong first-order electroweak phase transition is therefore potentially detectable by a future space-based interferometer~\cite{Caprini:2015zlo}.

\section{Taiji Program}
Due to geophysical noise there is no hope of detecting GWs at low frequencies from $10^{-4}$ to $1$ Hz
using ground-based interferometers.
The only way to detect these GWs is to put the GW detector in the quiet environment of space.
Such an instrument was put forward jointly by the European Space Agency and NASA
in the 1990s, i.e., the Laser Interferometer Space Antenna (LISA)~\cite{Danzmann:1997hm}.
Chinese scientists began to make proposals for space-based GW detection in the 2000s.
Recently, the Chinese Academy of Sciences has set up a strategic priority research
program that includes the pre-study of Taiji Program, a Chinese space-based GW detector~\cite{Hu:2017cbj,Cyranoski:2016gw,Gong:2014mca}.

Like LISA, Taiji consists of an equilateral triangle of three spacecrafts in orbit around the Sun.
Lasers are sent both ways between each pair of spacecrafts, and the differences
in the phase changes between the transmitted and received lasers at each spacecraft are measured.
Each spacecraft thus generates two interference data streams.
From these signals, we can construct the time variations of the armlengths
and then find GW signals.

\begin{figure}\small
  \centering
  \includegraphics[width=3.2in]{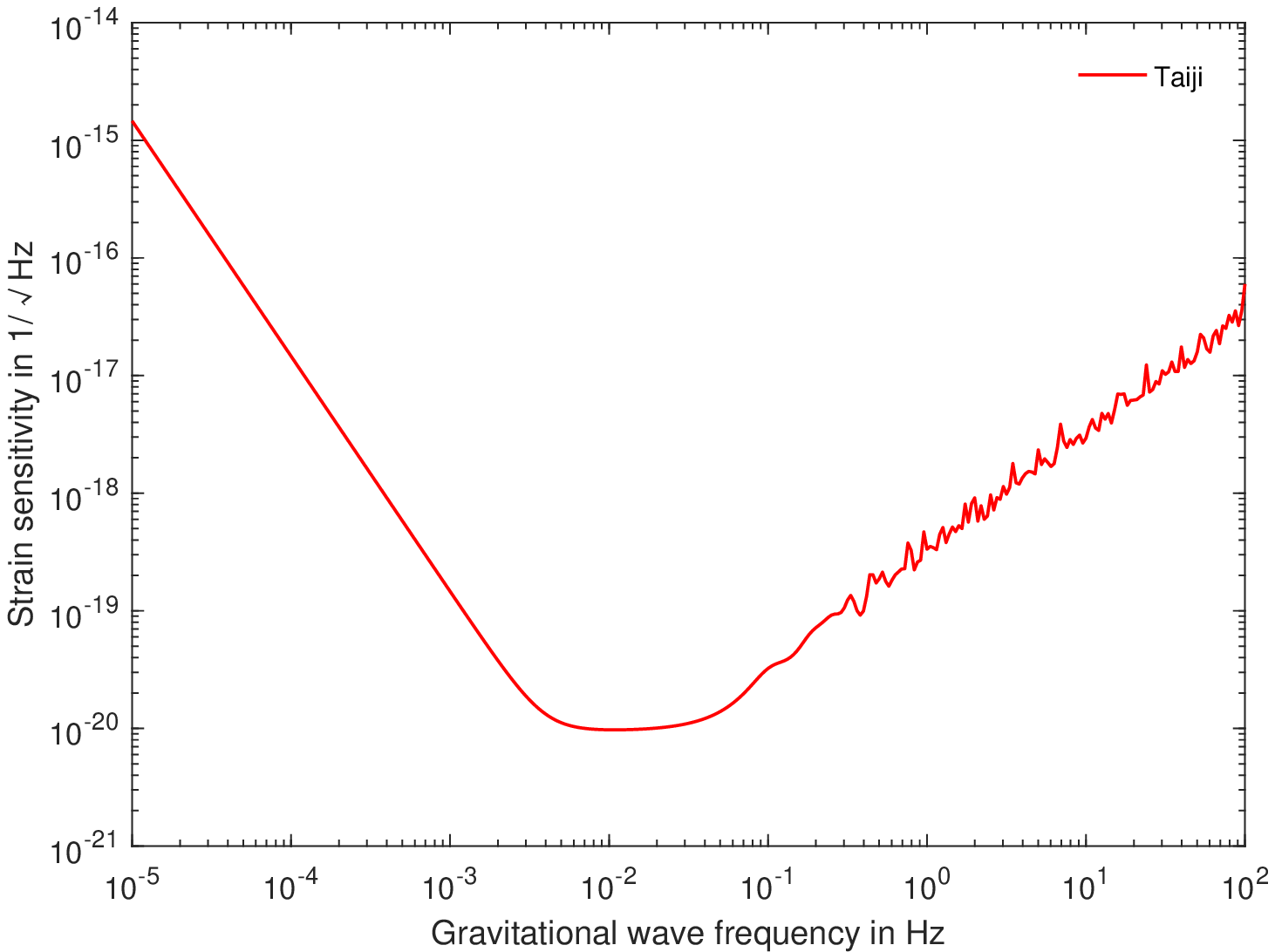}
  \includegraphics[width=3.2in]{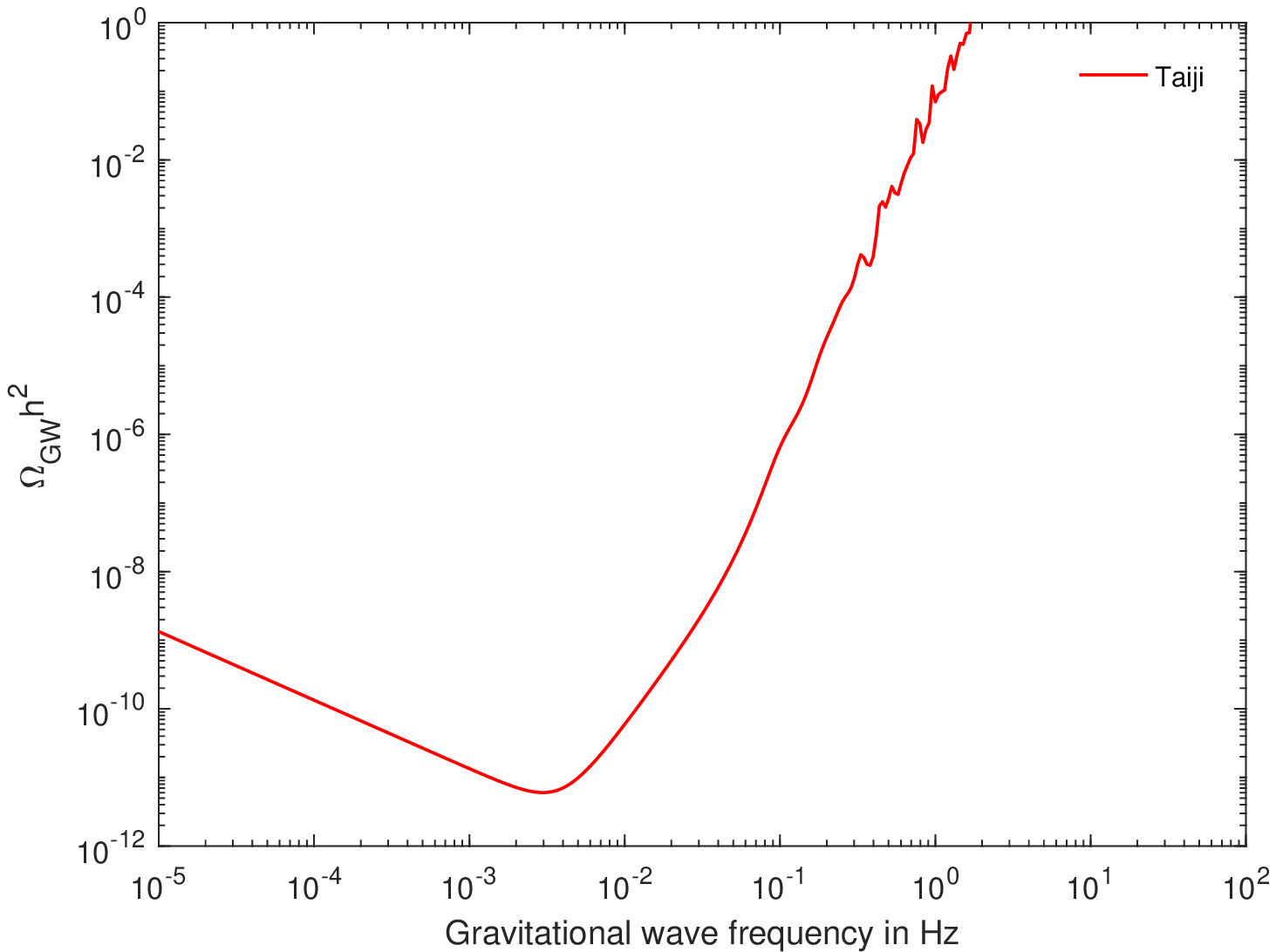}
  \caption{Strain sensitivity (left) and energy density (right)
  for Taiji with the spacecraft separation $3\times10^9$ m,
  telescope diameter $0.4$ m, laser power $2$ W, position noise $8$ pm Hz$^{-1/2}$,
  and acceleration noise $3\times 10^{-15}$ m s$^{-2}$ Hz$^{-1/2}$.}
  \label{fig:sensitivity}
\end{figure}

In Fig.~\ref{fig:sensitivity} we plot the strain sensitivity (left panel) and energy density (right panel) for Taiji
using the GW observatory designer tool described in~\cite{Barke:2014lsa}.
With the sensitivity curve~\footnote{for the strain sensitivity of Taiji please send an email to guozk@itp.ac.cn},
Taiji will be able to observe GW signals from several thousand of compact binaries in our galaxy,
and to discovery few tens of MBHB merges and EMRIs/IMRIs per year.
The SGWBs produced in the early Universe are expected to be detected by Taiji.

MBHBs are prime targets for space-based GW observations such as LISA and Taiji.
Using the Fisher information matrix approach,
we estimate how well Taiji will be able to measure MBHB parameters,
including the chirp mass $M_c=\eta^{3/5}M$,
systematic mass ratio $\eta=m_1 m_2/M^2$,
luminosity distance to the source $d_L$
and sky position of the binary, where $M=m_1+m_2$ is the total mass of the binary.
The error of binary's sky position in solid angle is given by
\ba
\Delta \Omega_s = 2\pi |\sin \theta_s|\sqrt{\langle \Delta \theta_s^2 \rangle \langle \Delta \phi_s^2 \rangle -\langle \Delta \theta \Delta \phi \rangle^2},
\ea
where $\theta_s$, $\phi_s$ are the colatitude and longitude of the binary in a polar coordinate system.
Table~\ref{tab:error} shows the results for equal-mass MBHBs with total intrinsic masses of $10^5 M_{\odot}$,
$10^6 M_{\odot}$ and $10^7 M_{\odot}$ at the redshift of $z = 1$.
We assume a flat $\Lambda$CDM cosmology with $\Omega_m=0.31$, $\Omega_\Lambda=0.69$, and $H_0=67.74$ km s$^{-1}$ Mpc$^{-1}$~\cite{Ade:2015xua}.
The observed mass $M$ is related to the intrinsic mass by the relation $M = (1+z)M_{\rm int}$.
We use the post-Newtonian waveform with a upper cutoff frequency of $f_{\rm isco} = c^3/6\sqrt{6}\pi GM$.
Moreover, the binary inclination angle is chosen to be $\iota = 45^{\circ}$.
We can see that the measurement accuracy are comparable with those of LISA~\cite{Lang:2007ge}.
\begin{table}[htbp]
  \centering
  \begin{tabular}{p{2cm} p{2cm} p{2cm} p{2cm} p{2cm}}
    \hline\hline
    Mass $(M_\odot)$ & $\Delta M_c / M_c$ & $\Delta \eta$ & $\Delta d_L /d_L$ & $\Delta \Omega_s \, ({\rm deg}^2)$ \\ \hline
    $10^5$ & $0.00000548$ & $0.0000627$ &  0.0409 &  6.11 \\
    $10^6$ & $0.0000287$  & $0.000141$  &  0.0471 &  7.66 \\
    $10^7$ & $0.000399$   & $0.00118$   &  0.0706 &  14.9\\
    \hline\hline
  \end{tabular}
\caption{Estimation of the errors of the chirp mass $M_c$, systematic mass ratio $\eta$,
luminosity distance to the source $d_L$ and angular resolution
for equal-mass MBHBs with total intrinsic masses of $10^5 M_{\odot}$, $10^6 M_{\odot}$ and $10^7 M_{\odot}$ at the redshift of $z = 1$.}
  \label{tab:error}
\end{table}

\begin{acknowledgments}
This work is in part supported by the Strategic Priority Research Program of the Chinese Academy of Sciences,
Grant No. XDA15020701, Grant No. XDB23030100 and No. XDB21010100,
by the National Natural Science Foundation of China Grants No. 11690021 and No. 11690022,
and by the Key Research Program of Frontier Sciences of CAS.
\end{acknowledgments}

\end{document}